\documentclass{article}
\usepackage{amsfonts}
\newcommand{\C}{{\mathbb C}}

\newcommand{\Hil}{{\mathcal H}}
\newcommand{\un}{{\mathfrak u}}
\newcommand{\eq}[1]{{\rm Eq.~(\ref{#1})}}
\newcommand{\mbf}[1]{\mbox{\boldmath$#1$}}
\newcommand{\bsgm}[1]{\mbf{\sigma}_\mathsf{#1}}
\newcommand{\btau}[1]{\mbf{\tau}_\mathsf{#1}}
\newcommand{\e}{\mathbf e}
\newcommand{\Id}{{\mathbf 1}}
\newcommand{\Mat}[2]{\left(\begin{array}{#1}#2\end{array}\right)}

\title{Algebras and universal quantum computations with higher dimensional
systems}

\author{Alexander Yu.\ Vlasov\\
Federal Radiological Center (IRH),\\
197101, Mira Street 8, St.-Petersburg, Russia
}

\begin{document}
  \maketitle

\begin{abstract}
Here is discussed application of the Weyl pair to construction of universal
set of quantum gates for high-dimensional quantum system. An application of
Lie algebras (Hamiltonians) for construction of universal gates is revisited
first. It is shown next, how for quantum computation with qubits can be used
two-dimensional analog of this Cayley-Weyl matrix algebras, i.e. Clifford
algebras, and discussed well known applications to product operator formalism
in NMR, Jordan-Wigner construction in fermionic quantum computations.  It is
introduced universal set of quantum gates for higher dimensional system
(``qudit''), as some generalization of these models.
Finally it is briefly mentioned possible application of such algebraic
methods to design of quantum processors (programmable gates arrays) and
discussed generalization to quantum computation with continuous variables.
\end{abstract}



\section{INTRODUCTION}
\label{sect:intro}

The models discussed here only recently were introduced in theory of quantum
computation \cite{Gott98,GKP01,VlaUNt}
for description of $d$-dimensional quantum systems ($d > 2$), but
they have rather long history. Initially some special matrices were introduced
by Arthur Cayley more than 100 years ago \cite{Z}, and 30 years later similar
construction was used
by Hermann Weyl for building of fundament of quantum theory \cite{WeylGQM}.
It is so-called Weyl pair of matrices (or elements of infinite dimensional
operator algebra, last case is not considered here in details for simplicity
and only briefly
discussed due to applications to quantum computations with continuous
variables). The Weyl approach seems very important in theory of quantum
computation, because it is only way to write an equivalent of Heisenberg
relations for finite-dimensional systems used there. In addition algebra has
important technical applications.  It is widely used in quantum computation
community after reintroducing few years ago with new name ``generalized
Pauli algebra.''

\section{Universal Quantum Gates and Lie Algebras}

``Quantum gate'' is special name for arbitrary unitary transformation.
``Quantum networks'' \cite{DeuGate} may be constructed using many
quantum gates and corresponds to some unitary evolution, ``quantum
computation.''
It is reasonable to consider some sets of elements
that may be used for construction (or approximation) of {\em any} unitary
transformation.
It is similar with using universal set of logical gates (Boolean functions)
like {\sf AND}, {\sf NOT} in classical networks and theory of computations.

In many models of quantum computations an elementary system is
described by finite-dimensional Hilbert space, but more difficult continuous
case is also may be considered. Quantum gates act on some
composite system with Hilbert space described as tensor product
(see \eq{Hil2prod} and \eq{Hilprod} below). For schemes of quantum
networks each ``wire'' corresponds to such system and it corresponds to
exponential growth of dimension of space of states with number
of such ``quantum wires.''

In simplest case the Hilbert space of the elementary quantum system,
{\em qubit}, is two-dimensional. More general case with
higher-dimensional systems is also used sometimes \cite{Gott98} and
considered in present paper. For $n$ qubits Hilbert space has dimension
$2^n$ and most general quantum evolution may be described by
$2^n \times 2^n$ unitary matrix.

The problem of universality is existence of some set or family
of unitary matrices $U_\kappa$ with possibility of decomposition
of any unitary matrix $U \in {\rm U}(2^n)$
\begin{equation}
U = U_{\kappa_1} U_{\kappa_2} \cdots U_{\kappa_p}
\label{UnG}
\end{equation}
or approximation $U \approx U_{k_1} U_{k_2} \cdots U_{k_p}$
with necessary or arbitrary precision, if the set is finite
\mbox{$k = 1,\ldots,K$}.

It should be mentioned, that formally the same gate acting on different
``wires'' corresponds to different matrices, but the additional indexes are
omitted in \eq{UnG} in sake of simplicity. Situation may be more difficult
in real applications, then system may not be simply presented as tensor
product\footnote{For example in theory of {\em decoherence free subspaces},
it is more known, but rather specific example.}, but it is not considered
in present work.

For quantum computation with $n$ qubits the Hilbert space $\Hil$ has special
structure due to representation as tensor power of $n$ two-dimensional
spaces $\Hil_2$
\begin{equation}
 \Hil = \underbrace{\Hil_2\otimes\cdots\otimes\Hil_2}_n .
\label{Hil2prod}
\end{equation}

 Due to such structure any unitary transformation of
some subsystem with $m < n$ qubits, $U_{\rm sub} \in {\rm U}(2^m)$ can be
expressed as transformation that acts only on $m$ elements (indexes)
in tensor product with $n$ terms \eq{Hil2prod} used for representation
of the Hilbert space for $n$ qubits. Such transformations are called
$m$-qubit gates.

The constructions may be simply extended to higher-dimensional quantum
systems. Let $d>2$ is dimension of Hilbert space $\Hil_d$ for one
``qu$d$it'', then $n$ systems may be described by $d^n$-dimensional space
\begin{equation}
 \Hil = \underbrace{\Hil_d\otimes\cdots\otimes\Hil_d}_n ,
\label{Hilprod}
\end{equation}
and unitary group U$(d^n)$. Non-binary quantum $m$-gates correspond to
elements of U$(d^m)$, $m \leq n$.

An important result of quantum theory of computation is existence
of universal set with two-qubit quantum gates \cite{DV95,Gates95,DeuUn}.

\medskip

There is so-called {\em infinitesimal} approach to construction of universal
set of quantum gates. It was discussed in original papers
\cite{DV95,DeuUn,UnSim} and described here only briefly. From physical
point of view it can be considered as using Hamiltonian
approach with Hermitian operators $H_k$ (Hamiltonian), instead of
$S$-matrix-like approach with unitary gates (see also \eq{UnA} below).

The infinitesimal approach uses Lie algebra $\un(N)$ of Lie
group U$(N)$ of unitary matrices\footnote{Here $N=2^n$ for quantum
computation with $n$ qubits and $N=d^n$ for higher-dimensional case.}
together with finite set of elements $A_k$ of this algebra
represented as some anti-Hermitian matrices $A_k = -A_k^\dag$. It is
possible also to write $i A_k \equiv H_k$ for some Hermitian matrices
$H_k = H_k^\dag$. Quantum gate may be expressed as
\begin{equation}
U_k^\tau = \exp(-i H_k \tau)
\label{UnA}
\end{equation}
with some real parameter $\tau$. Now instead of universal set of elements
of Lie group U$(N)$ used in \eq{UnG} it is possible to work with elements
$A_k = -i H_k$ of Lie algebra $\un(N)$.

\noindent{\bf Lemma} If the elements $A_k$ generate full Lie algebra $\un(N)$
by commutators, then it is possible to use set of gates described by \eq{UnA}
as universal set of gates \cite{DV95,DeuUn,UnSim}.

Subtleties of transition from Lie algebra to Lie group of gates based on
\eq{UnA} is not discussed with more detail and may be found in original works%
 \cite{DeuGate,DV95,DeuUn,UnSim,VlaUCl,VlaUNt}. Description of specific
mathematical problems may be found also in Ref.~\cite{Wea00}.

In physical applications $H_k = i A_k$ correspond to Hamiltonian and real
parameter $\tau$, may be infinitesimal \cite{DV95}, irrational \cite{DeuUn}
or it is possible to consider \eq{UnA} as one-parametric family, where $\tau$
is time \cite{UnSim}.

\section{Clifford Algebras}

If there is some associative algebra $\mathcal A$, then it is possible to
introduce structure of Lie algebra by using commutators $[A,B] = AB-BA$. Due
to relation with universal set of gates discussed above, it is reasonable
to look for algebras with ``simple commutation law'' with purpose to have
appropriate tools for construction and description of universal quantum gates.

As an example of such a commutation law it is possible to mention
\begin{equation}
 AB = \zeta BA \qquad(A,B \in \mathcal A,~~\zeta \in \C)
\label{zcom}
\end{equation}
It is commutative algebra for $\zeta = 1$ and anticommutative one for
$\zeta = -1$. It is useful also to consider more general case, then
$\zeta$ is arbitrary complex number and not all elements of algebra
satisfy equation \eq{zcom}, but only {\em different generators} of
the algebra.

Well known examples for $\zeta=-1$ are algebra of all $2 \times 2$ complex
matrices with generators are Pauli matrices\footnote{Formally only two
Pauli matrices are enough. Here is some subtlety, because 3D {\em real}
Clifford algebra uses all three Pauli matrices as generators and it is not
some pure mathematical trick, because may be used in description of real NMR
experiments \cite{SCH98,HD0}.}
\begin{equation}
 \bsgm x = \Mat{rr}{0&1\\1&0},\quad
 \bsgm y = \Mat{rr}{0&-i\\i&0},\quad
 \bsgm z = \Mat{rr}{1&0\\0&-1}
\label{PauliMat}
\end{equation}
and algebra of all $4 \times 4$ complex matrices with generators are four
Dirac matrices
\begin{equation}
\gamma_0=\Mat{rr}{0&\Id\\ \Id&0}, \quad
\mbf{\gamma} = \Mat{rr}{0&-\mbf{\sigma}\\ \mbf{\sigma}&0}
\label{DiracMat}
\end{equation}
Both algebras are {\em Clifford algebras} \cite{ClDir,Post}
defined by (anti)commutation law
\begin{equation}
 \e_k \e_j + \e_j \e_k = 2\delta_{jk},
\label{eCl}
\end{equation}
where $\e_k$ are generators of the algebra.

Main property of Clifford algebras used in mathematical and physical
application is algebraic decomposition of square root
\begin{equation}
 \left(\sum\nolimits_k c_k \e_k\right)^2 = \sum\nolimits_k c^2_k,
 \qquad c_k \in \C.
\label{ClRoot}
\end{equation}
Really \eq{ClRoot} often used for definition of Clifford algebra
instead of \eq{eCl}. For example, the property is used for representation
of Dirac equation as square root of Klein-Gordon equation. On the other
hand, \eq{ClRoot} seems not related {\em directly} with applications of
Clifford algebras in quantum computations discussed in present paper.

The Clifford algebras with even number of generators $2n$ are especially
convenient for description of quantum computation with $n$ qubits
due to so-called {\em product operator formalism} \cite{BR94,SCH98,HD0}
and some other applications \cite{VlaTmr,VlaUCl,VlaAGA}. It is enough
to look on standard presentation for generators of such algebra
\cite{ClDir,WeylGQM}
\begin{eqnarray}
 \e_{2k} & = &
 i {\,\underbrace{\Id\otimes\cdots\otimes\Id}_{n-k-1}\,}\otimes
 \bsgm x\otimes\underbrace{\bsgm z\otimes\cdots\otimes\bsgm z}_k \, ,
 \nonumber\\
 \e_{2k+1} & = &
 i {\,\underbrace{\Id\otimes\cdots\otimes\Id}_{n-k-1}\,}\otimes
 \bsgm y\otimes\underbrace{\bsgm z\otimes\cdots\otimes\bsgm z}_k \, ,
 \label{defE}
\end{eqnarray}
where $k = 0,\ldots,n-1$ and $\Id$ is unit matrix.

There are $2^{2n}$ different products of $2n$ generators $\e_k$ \eq{defE} and
it corresponds to presentation of the Clifford algebra via $2^n \times 2^n$
complex matrices. Such representation is useful both for quantum
$n$-gates as elements of Lie group U$(2^n)$ and Hamiltonians as
elements of Lie algebra $\un(2^n)$ of that group.

\smallskip

Jordan-Wigner representation
of fermionic {\em creation} and {\em annihilation} operators via
\begin{equation}
a_k^\dag = \frac{1}{2}(\e_{2k} + i \e_{2k+1}), \qquad
a_k = \frac{1}{2}(\e_{2k} - i \e_{2k+1})
\label{aferm}
\end{equation}
with standard properties
\begin{equation}
 \{a_k,a_j\} = \{a_k^\dag,a_j^\dag\} = 0, \qquad
 \{a_k^\dag,a_j\} = \delta_{kj}
\label{fermcom}
\end{equation}
provides useful links with fermionic quantum computation
\cite{WeylGQM,VlaTmr,BK00}\footnote{In Ref.~\cite{BK00}
were considered some points relevant to present consideration, but
terms ``local'' and ``spinless fermion'' seem not very appropriate.
For example, fermionic commutation relations \eq{fermcom} have
representation with two-matrices and it corresponds to
two-component wave vectors of particle with spin-1/2, so spin
may not be ``omitted.'' It is also not a model of spatially localized
particles. If coordinate part of wave function is taken into account,
only particles with definite momentum satisfy simpler expression like
\eq{fermcom} and it corresponds to maximal ``spreading,'' $\exp(ikx)$.}.

It should be mentioned, that an approach with creation and annihilation
operators was used for description of universal quantum simulators and
computers already in first works of Feynman \cite{FeySim,FeyComp}, but
the question again has some subtleties, because for description of
quantum computation it is also possible to use other operators produced
via \eq{aferm} from slightly changed $\e_k$ with $\Id$ instead of any
$\bsgm z$ in tensor product \eq{defE}. Sometime they are called now
{\em parafermionic} operators, but the special term may be a bit
misjudged, because such operators even simpler and more natural in formal
theory of quantum computation, than ``usual'' fermionic operators. Formally,
both kind of operators may be used equally in most constructions with
creation and annihilation operators used in Refs.~\cite{FeySim,FeyComp}.

\medskip

Let us return to theory of universal quantum gates. It can be
shown, \cite{VlaUCl} that the {\em commutators} of elements \eq{defE} generate
only $(2n^2+n)$-dimensional space and so set of gates based on these elements
via \eq{UnA} produces only $(2n^2+n)$D subgroup in $2^{2n}$D group of
all unitary gates \cite{VlaUCl}. It is interesting counter-example of
widespread belief about ``omnipresent universality.'' Anyway, appending
of only one element of third order like $i\e_0\e_1\e_2$ produces
universal set of gates \cite{VlaUCl}. For fermionic quantum computation
sometime is used representation with even number of operators \eq{aferm}
and so the result is in agreement with necessity of elements of second and
{\em fourth} order for universality in such a case \cite{BK00}.

The representation \eq{defE} is good for general algebraic approach or
fermions, but in usual theory of quantum computation more useful elements
have minimal amount of non-unit terms in tensor product. Formally
\eq{defE} produces universal set of $k$-gates with $k=1,\ldots,n$, but
it is very simple to construct universal set of two-gates.
It is enough to consider elements of second order $\e_k\e_{k+1}$,
because any such element is one- or two-gate. So universal set of
$2n+1$ two-gates may be based on following elements of Lie algebra
$\un(2^n)$ (Hamiltonians) \cite{VlaUCl}
\begin{equation}
 \e_0,
\quad \e_{k,k+1} \equiv i\e_k\e_{k+1}~~~(k = 0,\ldots,n-2),
\quad \e_{012} \equiv i\e_0\e_1\e_2.
\end{equation}
It may be checked, the last element of third order corresponds
to one-gate.

\section{Generalization with Weyl Pair}

The construction with Clifford algebra described above may not be
applied directly for non-binary quantum circuits, when each system
described by $d$-dimensional Hilbert space, $d>2$.
Hilbert space of composite system with $n$ qu$d$its is $d^n$-dimensional
and gates belong to U$(d^n)$.
But even in such a case the general principles are
very close \cite{VlaUNt}.

First, let us consider instead of Pauli matrices Weyl pair
\cite{WeylGQM,ConnesNG} described by \eq{zcom}
with $\zeta = \exp(2 \pi i / d)$:
\begin{equation}
 U V = \exp(2 \pi i / d) V U,\quad V V^\dag = U U^\dag = \Id,
\label{UVcom}
\end{equation}
where $U$ and $V$ are $d \times d$ complex matrices \cite{WeylGQM}
\begin{equation}
 U = \Mat{ccccc}{0&1&0&\ldots&0\\0&0&1&\ldots&0\\
 \vdots&\vdots&\vdots&\ddots&\vdots\\0&0&0&\ldots&1\\1&0&0&\ldots&0}\!,
 \quad
 V = \Mat{ccccc}{1&0&0&\ldots&0\\0&\zeta&0&\ldots&0 \\
 0&0&\zeta^2&\ldots&0 \\ \vdots&\vdots&\vdots&\ddots&\vdots\\
 0&0&0&\ldots&\zeta^{d-1}}\!.
\label{WeylPair}
\end{equation}
Lately it is called sometime ``generalized Pauli group'' \cite{Gott98}.

It is possible to consider three elements
\begin{equation}
\btau x = U, \quad
\btau y = \zeta^{(d-1)/2}U V, \quad
\btau z = V,
\end{equation}
with properties \cite{VlaUNt}
\begin{equation}
 \btau x \btau y = \zeta \btau y \btau x,\quad
 \btau y \btau z = \zeta \btau z \btau y,\quad
 \btau x \btau z = \zeta \btau z \btau x,\quad
 \btau\mu^d = \Id.
\label{taucom}
\end{equation}

Let us introduce set of complex $d^n \times d^n$ matrices similar with
construction \eq{defE} above:
\begin{eqnarray}
 {\mathbf t}_{2k} & = &
  {\underbrace{\Id\otimes\cdots\otimes\Id}_{n-k-1}\,}\otimes
 \btau x\otimes\underbrace{\btau z\otimes\cdots\otimes\btau z}_k \, ,
 \nonumber\\
 {\mathbf t}_{2k+1} & = &
  {\underbrace{\Id\otimes\cdots\otimes\Id}_{n-k-1}\,}\otimes
 \btau y\otimes\underbrace{\btau z\otimes\cdots\otimes\btau z}_k \, ,
 \label{defT}
\end{eqnarray}
where $k = 0,\ldots,n-1$.

The $2n$ elements ${\bf t}_k$ itself meet to analogue of commutation law
\eq{zcom}:
\begin{equation}
{\bf t}_j {\bf t}_k = \zeta {\bf t}_k {\bf t}_j ~~ (j < k),
\quad ({\bf t}_k)^d = \Id.
\label{TorDef}
\end{equation}
As it was suggested earlier, the rule makes test of universality simpler and
the elements are really useful for construction of universal quantum gates
for higher dimensional case ($d>2$) \cite{VlaUNt}.

It is interesting, an analogue of \eq{ClRoot} is also valid \cite{QFerm}
\begin{equation}
 \left(\sum\nolimits_k c_k \mathbf t_k\right)^d = \sum\nolimits_k c^d_k,
 \qquad c_k \in \C.
\label{dRoot}
\end{equation}
So the elements may be used for expression of algebraic $d$-th root,
but it is not relevant to present work.

The elements \eq{defT} are not Hermitian, but for construction
of universal gates via \eq{UnA} it is possible to use \cite{VlaUNt}
\begin{equation}
 {\bf t}^+_k = i ({\bf t}_k + {\bf t}^\dag_k), \quad
{\bf t}^-_k = ({\bf t}_k - {\bf t}^\dag_k).
\label{Hpm}
\end{equation}
Commutators of the elements generate whole Lie algebra and so
exponents \eq{UnA} of elements \eq{Hpm} may be used as universal
set of quantum gates \cite{VlaUNt}.

Formally initial expressions \eq{defT} represent $k$-gates
($k = 1,\ldots,n$), but it is again possible to use pairwise products
of the elements
\begin{equation}
 {\bf t}_0,\quad
 {\bf t}_{k,k+1} \equiv {\bf t}_k{\bf t}_{k+1}~~~(k = 0,\ldots,n-2)
\end{equation}
to construct universal set of two-gates. It is also necessary
to use generators \eq{UnA} with Hermitian operators expressed as sums
\begin{equation}
 {\bf t}^+_{k,k+1} = i ({\bf t}_{k,k+1} + {\bf t}^\dag_{k,k+1}), \quad
 {\bf t}^-_{k,k+1} = ({\bf t}_{k,k+1} - {\bf t}^\dag_{k,k+1}).
\label{H2pm}
\end{equation}

\smallskip

It should be emphasized, that for higher-dimensional quantum systems with
$d>2$ it is enough only $2n$ elements \eq{defT} and $4n$ elements \eq{Hpm}
of first order and it is not necessary to add some generator of fourth or
third order for $d=2$ (qubits and fermionic quantum computation). Such point
may be essential, for example elements of high order have difficulties of
implementation for fermionic quantum computation \cite{BK00} and even in
more general case such elements may have specific properties \cite{VlaUCl}.

If to recall that without such additional element quantum gates generate
only state of space with quadratic dimension (``complexity'') instead
of exponential \cite{BK00,VlaUCl}, then it is possible to suggest, that
for $d=2$ the question about additional element of third order may be
essential from point of view of universality, correspondence between
classical and quantum complexity, {\em etc.}. So absence of such subtleties
with one additional element for higher-dimensional system $d>2$ maybe
not so formal, as it could be suggested at first sight.

\smallskip

It was mentioned earlier relations of different standard techniques with
theory of quantum computation: unitary gates (S-matrix), Hamiltonian,
occupation numbers (annihilation and creation operators). The approach
discussed in present section has relation with yet another standard model,
{\em Weyl quantization} \cite{WeylGQM}. The Weyl pair $U,V$ \eq{WeylPair}
has many other applications in theory of quantum computations
\cite{Gott98,GKP01}, but specific new terminology 
used in some papers sometimes hides it.

The \eq{UVcom} is {\em Weyl representation of Heisenberg commutation
relation}. The Weyl approach may be even more significant in area
or research under consideration, because Heisenberg relation itself
may not be satisfied in finite-dimensional Hilbert spaces used in many
applications of quantum computations.

\section{Universal Quantum Processors}

Formally, the quantum network approach used above (and in many other papers)
resembles classical situation, then some chosen set of logical gates
(``chips'') are fastening on some board to build networks with
necessary functions, on the other hand, usual electronic computers have
fixed structure and different functioning is ensured by programmability.

An analogue of such classical computer is programmable quantum gate arrays
\cite{NC97,VMC01} or, simply, quantum processors \cite{VlaCla,VlaUQP,HBZ01,QP}.
Here are exists two different designs ``stochastic'' \cite{NC97,HBZ01}
and ``deterministic'' \cite{NC97,VMC01,VlaCla,VlaUQP,QP}.
Formally ``deterministic'' and ``stochastic'' processors initially have very
different design and only recently was found possibility of some confluence in
continuous limit \cite{QP}.

Anyway only first kind of processors will be
discussed in present paper, so word ``deterministic'' is omitted further
for simplicity.
Such quantum processor uses some version of ``controlled'' or ``conditional''
quantum dynamics \cite{QCtrl}. In such approach some quantum ``wire(s)'' are
used for encoding of operations on second set of ``wires.''
Quantum processor is special network with quantum system represented as
tensor product of two subsystem: ``program
bus'' and ``data bus,'' but there is some specific difficulties, because
different states of ``program bus'' should be orthogonal \cite{NC97} and
so in some meaning all valid programs are ``pseudo-classical,'' i.e.,
a superposition may not be used and would produce ``malfunction''
of quantum processor (entanglement of data and program, more details may be
found in original works \cite{NC97,VMC01,VlaCla,VlaUQP,HBZ01,QP}).

Let us consider a particular design
of quantum processors~\cite{VlaCla,VlaUQP}. There are three buses:
{\em quantum data bus}, there any superposition are possible,
{\em intermediate bus} used for choice of particular quantum
gate applied to data bus on current state. Intermediate bus is also
some quantum system, but only fixed number of orthogonal basic states
may be used to produce proper work of quantum processor. And, finally,
there is (pseudo-)classical bus necessary to programming of changes in
intermediate bus for performing necessary sequence of quantum gates on
quantum data bus.

In simplest case the pseudo-classical bus is simply ``cyclic ROM,'' but
sequences of universal gates used for implementation of some quantum
algorithms may be too long for realistic implementations
using the ROM based approach. So it is better,
if sequence of gates may be described by some simple law and processor may
use some program of minimal size, i.e. it should be set of instructions
like
\begin{center}
{\tt repeat \{ repeat $U_5$ 10 times; repeat $U_7$ 20 times \} 3 times},
\end{center}
but these issues related with application of {\em reversible} classical
programming are out of scope of present work.

It is clear, that theory of (deterministic) universal quantum processors
resembles theory of universal quantum networks, but sequence of quantum gates
is not implemented statically, and each next operation is dynamically
chosen on each step of evolution, accordingly with program.

At present time a statement, that some particular set of gates is universal,
has limited significance, because property of universality is already proved
for very general sets of quantum gates. Discussed algebraic approach is
important due to constructivity, i.e., there are described concrete
and not very difficult algorithms \cite{VlaUCl,VlaUNt} for decomposition
or approximation of arbitrary unitary operators.

It may be not very actual for some standard quantum algorithms, when
construction of quantum networks is already developed
and optimized during many years in works of different researchers, but
{\em universality} of quantum processors suggests existence of algorithm
for decomposition of arbitrary unitary matrix with given precision.
Maybe realistic applications are related with mixed approach, when
the universal algorithm of decomposition is used for $k$-gates with
not very big $k$ (i.e., $d^k$ has appropriate order) and the gates are used
as building blocks for more difficult quantum algorithms.

\section{Continuous Case}

The \eq{UVcom} is {\em Weyl representation of Heisenberg commutation
relations} and valid both in finite and continuous (infinite) case
\cite{WeylGQM}. For $d \to \infty$ two families of operators with property
\eq{zcom} may be expressed as
\begin{equation}
 U_a \colon \psi(q) \mapsto \psi(q+a), \quad
 V_b \colon \psi(q) \mapsto e^{i b q} \psi(q),
\label{UaVb}
\end{equation}
where $a,b$ are two real parameters and $\zeta = \exp(iab)$ in \eq{zcom}.
It is possible also to write the operators in {\em exponential form}
\begin{equation}
 U_a = e^{i a \mathbf p}, \quad V_b = e^{i b \mathbf q},
\label{expUaVb}
\end{equation}
where {\bf p}, {\bf q} are operators of coordinate and momentum
\cite{WeylGQM}.

All necessary construction used for finite $d$ above may be generalized
on continuous case \cite{QFerm}, but such universal gates for continuous
case have ``bi-exponential'' structure, like
\begin{eqnarray*}
& e^{(U_a-U^\dag_a)\tau}=
 \exp\bigl([\exp(i a \mathbf p)- \exp(-i a \mathbf p)]\tau\bigr),
\\
& e^{i(U_a+U^\dag_a)\tau}=
 \exp\bigl(i[\exp(i a\mathbf p)+\exp(-i a\mathbf p)]\tau\bigr).
\end{eqnarray*}

Such approach not only produces universal set of gates for one quantum
continuous variable like in some other works \cite{LlBr98}, but for
arbitrary amount of such variables. It is only necessary instead of
${\bf t}_k$ \eq{TorDef} in all expressions uses their continuous analogues
described in Ref.~\cite{QFerm}. All constructions are based on
straightforward change of $U$, $V$ \eq{WeylPair} to $U_a$, $V_b$ \eq{UaVb}.
Here infinite-dimensional Hilbert space is represented as space of
complex function with $n$ real variables.

Most difficult part of proof used in Ref.~\cite{VlaUNt} related with
consideration of different nonstandard cases with $\zeta^k=1$, but it
becomes trivial in infinite-dimensional case, because it is possible to
choose  $\zeta^k \ne 1$ for any integer $k$.
Here is again possible to construct set of universal two-gates for arbitrary
{\em finite} number of continuous quantum variables.

Physical meaning may depend on choice of particular conjugated
variables {\bf p}, {\bf q}. For example, if {\bf q} is usual coordinate,
expressions
$$
e^{(V_b-V^\dag_b)\tau}=
 \exp\bigl(-i\sin(b\, x)\tau\bigr),
\quad
e^{i(V_b+V^\dag_b)\tau}=
 \exp\bigl(i\cos(b\, x)\tau\bigr),
$$
look more realistic as Hamiltonian for particle in periodic potentials,
but more difficult analogues of operators \eq{H2pm} maybe look more
strange. On the other hand, the continuous generalizations of $\mathbf t_k$
are based on some non-canonical form of Weyl commutation relations \cite{QFerm}
and so may be relevant to description of real systems via Weyl quantization
(and analogues of Wigner functions).




\begin{thebibliography}{99}
 \bibitem{Gott98} D. Gottesman, ``Fault-tolerant quantum computation
  with higher-dimensional systems,'' {\em Lect. Not. Comp. Sci. \bf 1509},
  302 (1999); {\em Preprint} {\tt quant-ph/9802007}.
 \bibitem{GKP01} D. Gottesman, A. Kitaev, J. Preskill, ``Encoding qubit
  (qudit) in an oscillator,'' {\em Phys. Rev. A \bf 64}, 012310 (2001);
  {\em Preprint} {\tt quant-ph/0008040}.
 \bibitem{VlaUNt} A. Yu. Vlasov,  ``Noncommutative tori and universal
  sets of non-binary quantum gates,''
  {\em Preprint} {\tt quant-ph/0012009} (2000);
  {\em J. Math. Phys.} {\bf 43}, 2959--2964 (2002).
 \bibitem{Z} C. Zachos, private communications.
 \bibitem{WeylGQM} H. Weyl, {\em The Theory of Groups and Quantum
  Mechanics}, (Dover Publications, New York 1931).
 \bibitem{DeuGate} D. Deutsch, ``Quantum computational networks,''
  {\em Proc. R. Soc. London Ser. A \bf 425}, 73--90 (1989).
 \bibitem{DV95} D. P. DiVincenzo,
 ``Two-bit gates are universal for quantum computation,''
  {\em Phys. Rev. A \bf 51}, 1015--1022 (1995).
 \bibitem{Gates95} A. Barenco, C. H. Bennett, R. Cleve, D. P. DiVincenzo,
  N. Margolus, P. W. Shor, T. Sleator, J. A. Smolin, and H. Weinfurter,
 ``Elementary gates for quantum computation,''
 {\em Phys. Rev. A \bf 52}, 3457--3467 (1995).
 \bibitem{DeuUn} D. Deutsch, A. Barenco, and A. Ekert,
 ``Universality in quantum computation,''
  {\em Proc. R. Soc. London Ser. A \bf 449}, 669--677 (1995).
 \bibitem{UnSim} S. Lloyd, ``Universal quantum simulators,''
  {\em Science \bf 273}, 1073--1078 (1996).
 \bibitem{Wea00} N. Weaver, ``On the universality of almost every
  quantum logic gate,'' {\em J. Math. Phys.} {\bf 41}, 240--243 (2000).
 \bibitem{ClDir} J. E. Gilbert and M. A. M. Murray, {\em Clifford algebras and
  Dirac operators in harmonic analysis}, (Cambridge University Press,
  Cambridge 1991).
 \bibitem{Post} M. M. Postnikov, {\em Lie groups and Lie algebras},
  (Nauka, Moscow 1982).
 \bibitem{ConnesNG} A. Connes, {\em Noncommutative Geometry},
  (Academic Press, San Diego 1994).
 \bibitem{FeySim} R. P. Feynman, ``Simulating Physics with
  Computers,'' {\em Int. J. Theor. Phys. \bf 21}, 467--488 (1982).
 \bibitem{FeyComp} R. P. Feynman, ``Quantum-Mechanical
  Computers,'' {\em Found. Phys. \bf 16}, 507--531 (1986).
  \bibitem{BR94} B. Boulat and M. Rance, ``Algebraic formulation of the
  product operator formalism in the numerical simulation of the dynamic
  behaviour of multispin systems,'' {\em Mol. Phys. \bf 83} 1021--1039 (1994).
 \bibitem{SCH98} S. S. Somaroo, D. G. Cory, T. F. Havel, ``Expressing the
  operations of quantum computing in multiparticle geometric algebra,''
  {\em Phys. Lett. \bf A 240} 1--7 (1998).
 \bibitem{HD0} T. F. Havel, C. J. L. Doran, ``Geometric algebra
  in quantum information processing,''  {\em Preprint}
  {\tt quant-ph/0004031} (2000).
 \bibitem{VlaTmr} A. Yu. Vlasov, ``Quantum gates and Clifford algebras,''
  {\em Preprint} {\tt quant-ph/9907079}, (1999).
 \bibitem{BK00} S. Bravyi, A. Kitaev,``Fermionic quantum computation,''
  {\em Preprint} {\tt quant-ph/0003137} (2000).
 \bibitem{VlaUCl} A. Yu. Vlasov,  ``Clifford algebras and universal
  sets of quantum gates,''
  {\em Preprint} {\tt quant-ph/0010071} (2000);
  {\em Phys. Rev. A \bf 63}, 054302 (2001).
 \bibitem{VlaAGA} A. Yu. Vlasov,``Clifford algebras, noncommutative tori
  and universal quantum computers,''  {\em Preprint} {\tt quant-ph/0109010}
  (2001).
 \bibitem{QFerm} A. Yu. Vlasov,``Notes on Weyl--Clifford algebras,''
 {\em Preprint} {\tt math-ph/0112049} (2001).
 \bibitem{NC97} M. A. Nielsen, I. L Chuang, ``Programmable quantum
 gate arrays,'' {\em Phys. Rev. Lett. \bf 79} 321--324, (1997).
 \bibitem{VMC01} G. Vidal, L. Masanes, J. I. Cirac, ``Storing
 quantum dynamics in quantum states: stochastic programmable gate for
 $U(1)$ operations,'' {\em Preprint} {\tt quant-ph/0102037}, (2001);
 {\em Phys. Rev. A \bf 88}, 047905 (2002).
 \bibitem{VlaCla} A. Yu. Vlasov, ``Classical programmability is enough
  for quantum circuits universality in approximate sense,''
  {\em Preprint} {\tt quant-ph/0103119} (2001).
 \bibitem{VlaUQP} A. Yu. Vlasov, ``Universal quantum processors
  with arbitrary radix $n \ge 2$,''
  {\em Preprint} {\tt quant-ph/0103127} (2001).
 \bibitem{HBZ01} M. Hillery, V. Buzek, M. Ziman, ``Probabilistic
 implementation of universal quantum processors,''
 {\em Preprint} {\tt quant-ph/0106088}, (2001);
 {\em Phys. Rev. A \bf 65}, 022301 (2002).
 \bibitem{QP}  A. Yu. Vlasov, ``ALEPH-QP: Universal Hybrid Quantum
  Processors,'' {\em Preprint} {\tt quant-ph/0205074} (2002).
 \bibitem{QCtrl} A. Barenco, D. Deutsch, A. K. Ekert, R. Jozsa,
 ``Conditional quantum dynamics and logic gates,''
 {\em Phys. Rev. Lett. \bf 74}, 4083--4086 (1995).
 \bibitem{LlBr98} S. Lloyd, S. L. Braunstein, ``Quantum computation
 over continuous variables,'' {\em Preprint} {\tt quant-ph/9810082}, (1998);
 {\em Phys. Rev. A \bf 82}, 1784--1787 (1999).
\end{thebibliography}
\end{document}